# Electoral Competition with Credible Promises and Strategic Voters

By

Shiladitya Kumar

Working Paper

# Extended Abstract

How can voters induce politicians to put forth more proximate (in terms of preference) as well as credible platforms (in terms of promise fulfillment) under repeated elections? Building on the work of Aragonès et al. (2007), I study how reputation and re-election concerns affect candidate behavior and its resultant effect on voters' beliefs and their consequent electoral decisions. I present a formal model where, instead of assuming voters to be naive, I tackle the question by completely characterizing a set of subgame perfect equilibria by introducing non-naive (or strategic) voting behavior in the mix. I find that non-naive voting behavior - via utilizing the candidate's reputation as an instrument of policy discipline post-election - aids in successfully inducing candidates to put forth their maximal incentive compatible promise (amongst a range of such credible promises) in equilibrium. Through the credible threat of punishment in the form of loss in reputation for all future elections – non-naive voters are able to gain a unanimous rise in their expected utility than when they behave naively. In fact, comparative statics show that candidates who are more likely to win are more likely to keep their promises. In such a framework, voters are not only able to better bargain for more credible promises but also end up raising their expected future payoffs in equilibrium. Including such forms of strategic behavior thus reduces cheap talk by creating a credible electoral system where candidates do as they say once elected. Later, I present an analysis by including *limited punishment* as a suitable political accountability mechanism into the framework.

# Motivation of the Formal Study [1]

## Introduction

Elections play a singularly important role in the determination of public policies in representative democracies. The presence of the electoral mechanism provides a way to choose individuals who not only act as representatives of a section of the electorate but also take decisions on their behalf, which the electorate hope will meet their economic needs. Thus, an election seeks to resolve persistent conflicts arising amongst a select group of competing majorities, which in turn help to convert voters' preferences into collective outcomes (Banks and Duggan, 2008). The study of the resultant attritional head to head often seen in these electoral competitions have led to the already burgeoning literature in the sphere of the political economy of public policy, where the gilt-edged tools of game theory have been steadily applied to study the underlying mechanisms at work.

Grounded in the traditional approach to model electoral behavior, Downs (1957) in his critical analysis directs our collective attention to a set of crucial assumptions, primary of them being that parties contest an election with the sole objective of winning it. That is, parties are office-motivated. The vital breakthrough of this assumption was that for a two-party election setup, with both the parties having the same set of information regarding voters' preferences, the equilibrium will be one of full convergence to the median voter's policy pref-erence. This result is the case when the policy space is one-dimensional, and voters have single-peaked preferences. We should also note in this regard that the above Downsian result is a unique best response strategy for both the parties at work. That is, the convergence to the median voter's policy is a Nash equilibrium. However, in the presence of *valence* on the part of the candidates, or when they have policy preferences, and there exists uncertainty regarding the position of the median voter, we will find such convergence disappearing in equilibrium.

---

[1]Shiladitya Kumar: Ph.D. Candidate ($6^{th}$ Year), Dept. of Political Science, University of Houston. Email: skumar33@cougarnet.uh.edu. I thank Shubhro Sarkar, Francisco Cantù, Jessica Gottlieb, Federica Izzo, Carlo Prato, Juan Dodyk, and participants at MPSA 2023, Stony Brook Game Theory Conference 2023, and the Comparative Politics and Formal Theory Conference 2023 for their feedback. All remaining errors are my own. This is a work in progress; please do not cite.

The benchmark model of Aragonès et al. (2007), upon which I have based my present study, directs our collective attention to a particular form of political campaigns which is referred to in the literature as *credible commitment*. It is, in fact, interesting to note that such campaign promises often act to solve a coordination problem. Such issues might arise in multi-principal agency models where campaign promises as these can provide a common focal point for voters to choose amongst multiple equilibria in a repeated game (Aragonès et al., 2007). In a way, a campaign promise is a mechanism that allows multiple principals to converge on a single point of a rule to control the actions of the agent in question (Aragonès et al., 2007). These announcements allow the voters to set their beliefs regarding the candidates seeking election and in a way to link them to a future voting belief during a run-up to re-election. The model (Aragonès et al., 2007) then sets upon to analyze the conditions under which candidates' reputations affect voters' beliefs over what policy shall be implemented by the winning candidate in sequential elections with complete information. Under such an ideological setup, the model then seeks to find out the range of maximal credible promises that will be believed by voters and honored by candidates in equilibrium.

In tune, I presently aim to extend the benchmark model in two different dimensions, which are -
(a). What would the maximal credible promise be, if instead of the presence of naive voters (Aragonès et al., 2007), we assume non - naive, strategic voting behavior in equilibrium? Under the premises of an infinitely repeated electoral game with complete information, Aragonès et al. (2007) found out that - (a). The value of the maximal credible promise[2] is zero for smaller values of the discount factor $\delta$.

(b). The value of $d^*$ gives us a fraction (between 0 and 1) for intermediate values of $\delta$.

(c). The value of $d^*$ is one for higher values of $\delta$.

The maximal range of credible distance, denoted by $d^*$ characterizes the set of promises which are *incentive-compatible* in the sense that candidates who have never reneged, fulfill all the promises they had made before each election had taken place, and which the voters (more importantly) believe, in equilibrium.

However, when the assumption of naive voting behavior is supplanted by one that of non - naive voting behavior, we find that -

(a). The value of $d^*_G$ (derived when a candidate's opponent is *good*, in the sense that he has never reneged on a promise - is a positive fraction (between 0 and 1) for intermediate values of the discount factor(given by $\delta$); while it is equal to 1 and 0 for sufficiently higher and lower values of $\delta$ respectively.

(b). The value of $d^*_B$ (derived when a candidate's opponent is *bad*, in the sense that he has reneged on a promise before - is a positive fraction (between 0 and 1) for all values of the discount factor(given by $\delta$)

(c). The value of $d^*_G$ lies below the Aragonès et al. (2007)'s result for small and intermediate values of $\delta$ but converges to the Aragonès et al. (2007)'s result for all values of $\delta \in [0.75,1]$

As would be examined in greater detail later in the present study, the assumption of *naive voting behavior* (Aragonès et al., 2007) not only allows the candidates to both put forth policy announcements that are less than the farthest possible incentive compatible promises (given particular scenarios described later) but also maintain their reputation as a non-reneging, *good* candidate for all future elections to come. Therefore, in many a such situations, the candidate is able to 'save up' on utility accruing from his policy implementation and his promise, at the expense of the voters. The *naive voter* fails to counter this *loss* in its own utility arising from the candidate not putting forth the closest possible (to the median voter) incentive compatible promise. In turn, the latter of the results can intuitively be thought of as - under *non - naive voting behavior*, the voters would make the more moderate (winning) candidate to make an even better (or more) credible promises than in the benchmark model or face the brunt of being labeled *bad* for the rest of all future elections (Aragonès et al., 2007). Thus, such 'coercive' behavior on part of the voters would reduce the worth of maintaining a promise in equilibrium, and would thus lead to a reduction in the maximal promise that can be sustained (Aragonès et al., 2007)[3]. Therefore, it is correct in suggesting that such strategic behavior on the part of the voters is a form of equilibrium selection.

---
[2]Henceforth to be cited as $d^*$

## Related Literature

Most of the work on the political economy of electoral competition has been based on the seminal study of Downs (1957), where voters chose amongst a set of announced policies by the candidates, which have been termed as Models of pre-election politics and separately as those of post-election politics (Klingelhöfer, 2015). These categories of political models have been propounded by the work of Ferejohn (1986), where the politicians are induced to put in more effort in their campaign trails by the voters. Such models of post-election have also been applied to issues about accountability, or to those regarding limits on rent extraction due to the possibility of losing out in future elections (if the politicians do not comply)(Barro, 1973). In such a framework, candidates are assumed to be both identical and have no policy preferences of their own, while being judged solely by their past performances rather than on campaign promises or cheap talk they might have taken refuge in.

One of the seminal studies based on the premise of the Downsian two-party system can be seen in the work by Alesina (1988), who via a model of dynamic inconsistency showed that in a one-shot election framework the parties do not arrive at a time-consistent converging policy platform. Under the assumptions of rational, forward-looking voters with parties who are office-motivated, he proved that the case of full convergence does not take place because now the parties care not only about getting elected to office but also about the quality of the policy to be implemented as well. Accordingly, the study by Dixit et al. (2000), shows that the incumbent does not move from his promised policy to his most preferred one in hope of not triggering "political compromise" with the challenger.

---

[3]Our results follows the intuition provided Aragonès et al. (2007) in page 866 of the section titled 'Discussion'.

A similar formal work (to the present thesis) has been done by Van Weelden (2013) - where the author shows that the best stationary subgame perfect equilibrium is the one where 'non-median candidates' are elected over others (from a heterogeneous pool of candidates) whose policy preferences lie closer to the median voter. The chosen candidate then decides to implement both his policy as well as the amount of rent-seeking to engage in, once elected to office (Van Weelden, 2013). Even though the above study looks at the potential divergence in policy implementation regarding rent-seeking activities, it does employ a similar approach to that of the benchmark model. The similarity lying in that both the frameworks end up integrating the concepts of political accountability and electoral competition.

Duggan (2000) later furthered the work of Alesina (1988) by incorporating elements of both retrospective as well as prospective voting behavior to find a symmetric, stationary equilibrium, where the representative voter is decisive in sequential elections setting. The basic premise is that, with a challenger drawn randomly from the electorate in each period to run against the incumbent, the latter wins the re-election if and only if his most recent policy platform gives the median-voter at least as much payoff as he would have received from the challenger. Another work of interest in this regards has been done by Banks and Duggan (2008), where they undertook an analysis consisting of a dynamic, multi-dimensional policy spectrum (excluding campaigns), and characterized a set of equilibria using strategies. Without considering prospective evaluations of candidates in each subsequent period of analysis, the incumbent faced a random opponent - where it was shown that an individual voter would vote for the incumbent whose proposed policy would lead to an exogenously determined threshold utility level for the said individual.

Furthermore, Duggan and Fey (2006) investigated the properties of an infinitely repeated game of electoral competition with complete information (without campaign credibility) and retrospective voting. Since the candidates are purely office- motivated, the authors duly noted that they are indifferent to the policies they would implement if elected. Aragonès et al. (2007) too developed a dynamic model of repeated elections with complete information, where candidates' reputations affect voter beliefs regarding the policies that may be implemented in an election by the winning candidate. One of the sets of interesting results was that in equilibrium the extent to which these promised policies are credible is an increasing function of the candidates' reputation. In the same vein, Grosser and Palfrey (2014) advanced the work by Aragonès et al. (2007) by allowing for the endogenous entry of citizens with private information about their respective bliss points. The authors used the citizen-candidate-entry model of Besley and Coate (1997) with the added assumption of polarized candidates bliss points, to come up with a unique symmetric equilibrium of the entry game. They further showed that whereas the more "extreme" candidate entered the contest, the more "moderate" candidate stayed away from the fray.

Later, the work of Klingelhöfer (2015) reconsidered the analysis by combining the idea of both forward-looking as well as backward-looking voters in the literature of electoral competition. Under both the assumptions of no uncertainty with regards to voters' policy preferences and commitment by interested candidates to a given policy platform - he showed that voters could limit the extent of rent collection by the elected politician with the aid of a purely backward-looking voting model. Meanwhile, the median voter policy is implemented in the standard forward-looking model of electoral competition, which the voters do so by carrying out a simple lexicographic voting strategy. In the same vein, Penn (2009) contributed to the literature on non - naive voting behavior in formal models of electoral competition by developing a bargaining model over continuing programs, in which, agents rank policies not only on the basis of the payoffs they derive today but also with regards to those alternatives they are likely to receive in future. The model infused both aspects of farsightedness as well as a continuation of policies to characterize dynamically stable voting equilibria of electoral behavior.

In contrast to the benchmark study, Panova (2017) models campaign promises as "pure cheap talk" (as also seen in Kartik and Van Weelden (2019)) to understand its influence over voters, in a two-period elections setup. Unlike Aragonès et al. (2007) where candidate preferences are public information, Panova (2017) characterizes the equilibrium conditions assuming candidate's policy preferences as private information. The voters (in the latter study) form their electoral decisions based on the signals put forth by the candidates in their election campaigns. In accordsance to these signals, the candidate is chosen who lies closer to the median voter's preferences. Voters are "retrospective" rather than "prospective" in this model. More recently, Born et al. (2018) analyzed the effect of campaign promises on electoral behavior but in an experimental setting, where the politicians make nonbinding promises regarding the proportion of splitting an endowment between themselves and the group. They found that the credibility of the promises affect both voters' beliefs and how they vote regarding how much the politician will end up contributing to the public fund - in the form of an inverted U-shaped curve - with lowered credibility relating itself to higher promises.

Interestingly, in a study closely aligned to the present thesis, Kartik et al. (2017) look into the possible effects of reputation (*good reputation effects* and *bad reputation effects*) on politician's re-election prospects. In contrast to the benchmark model which holds information regarding candidates' policy preferences as public information, the above work builds a dynamic model of electoral accountability under the assumption that candidate preferences are private knowledge. Through electoral promises, candidates aim to influence the beliefs of the voters in order to both build a sound reputation as well as enhance future electoral prospects. They find that, ceteris paribus, *good reputation* causes incumbency disadvantage when the incumbent has a stronger reputation than its newly-elected counterpart. While *bad reputation effects* causes incumbency advantage under the same setting.

# The Main Model and its Extensions

## The Benchmark model of Aragonès et al. (2007)

In infinitely repeated elections, promises are credible in equilibrium only if reputation has value (Aragonès et al., 2007). However, a promise can always be reneged upon if it is in the interest of the candidate to do so. This is because future payoffs coming out of maintaining such a promise may be different from those when a promise has been broken. Hence, promises such as these often alter the beliefs of the voters regarding the choices the candidates will make once elected. The voters understand that this results from the self-interest of a candidate to maintain his promise even in the presence of a short-term gain from reneging. Voters also do take into account that the threat of future punishment for reneging won't deter all promises from being broken. That is, if the promised policy platform is far enough from a candidate's ideal point with the gain from short-term reneging being sufficiently high, it is always in the self-interest of the candidate to renege on the promised platform. In short, the ability of the candidate(s) to alter voters' beliefs is not "technologically" given; rather it is an equilibrium phenomenon (Aragonès et al., 2007).

The model assumes complete information in the way that voters perfectly know candidates' preferences at the time they go to vote (Aragonès et al., 2007). It is further assumed that, at each election, voters view each of the candidates as either *good* or *bad* - this, in essence, means voters hold candidates who have reneged in one of the past elections as having a *bad* reputation, whereas candidates who have never reneged are held with having a *good* reputation. In fact, in the model voters believe that promises coming from a *good* candidate, as opposed to a one with a *bad* reputation. After the end of each election, a winning (*good*) candidate compares the utility he would have gained from maintaining his promise with that of the one-time benefit of reneging on any promise he may have made. However, it must be noted that candidates with a *bad* reputation will always implement their ideal point if and when they are elected to office.

An interesting point to note is the relationship between the values of the discount factor in equilibrium and to the extent to which a candidate's policy will be able to alter the beliefs of the voters (Aragonès et al., 2007). As long as the discount factor is large enough, the candidates will be able to alter the beliefs with regards to the platform they propose to undertake. Thus, if the expected value of getting elected is positive for all future periods, then the value of maintaining a *good* reputation tends to infinity as the discount factor tends to one. Hence, along with such high values of reputation which the candidates intend to hold on to, all promises in such a setting will be kept and in turn, believed by the voters.

**The Model**

Two candidates, L and R, compete for office in each of the repeated elections. The timing of events for each election is – (a). *Campaign Stage*: Each of the two candidates announces a platform simultaneously. They must now decide whether to make a promise about the policy that they will implement or send a message devoid of promises. (b). *Voting Stage*: Now, the voters receiving the messages from the candidates, decides to vote for the one whose policy platform maximizes their expected utility - depending entirely on the policy platform believed by the individual voters. (c). *Office Stage*: At the final stage of the game, the winner implements the policy.

The utility function assumed for both the agents in the model is represented as –

$$u_i(x) = -|x - x_i| \qquad (1)$$

Here, $x_i$ represents agent i's ideal point. The policy interval is given by $[-1, 1]$ with the ideal point of the median voter remaining the same for all subsequent elections, such that $x_m = 0$. With elections taking place over time, the voters vote for the candidate whose promised pol-icy choice is most preferred by them. Moreover, the discount factor considered in the model is $\delta \in [0, 1)$, which is held to constitute the weights the candidates' put on their future payoffs.

The model further assumes that the policy preferences of the candidates alter for each election, where the ideal point of candidate R denoted by $x_R \in [0, 1]$ is drawn randomly from a uniform distribution on the interval [0,1]. Similarly, the ideal point for candidate L denoted by $x_L$ also constitutes a random draw from the uniform distribution on the range [-1,0] (Aragonès et al., 2007). An important point to note under such a scenario is that each of these draws is independent of one another and also of all previous draws before each election takes place. Moreover, at the beginning of each period, both the candidates as well as the voters come to know about the ideal points of both the candidates as the candidates are of the median voter (Aragonès et al., 2007).

As part of the candidates' one-shot election strategy, they select a pair $(p,x)$, where $x \in [-1, 1]$ represents the policy to be implemented by the candidate if he wins the election and $p \in [-1, 1] \cup \Phi$ represents the candidates' announcements (promised or otherwise) before the *voting stage*. Thus, the voters knowing the candidates' ideal points, update their beliefs regarding a candidate's policy platforms (in case he wins the election) before going to the polling booth. These updates are based on the announcements made by the respective candidates at the *campaign stage*. Given the updated beliefs, the voters then decide to vote for that candidate who maximizes their expected utilities (Aragonès et al., 2007). A point to note in this regard is that voters believe that in the absence of promises, a candidate will choose his ideal point once elected to office. Thus, although campaign promises do not affect the flow of utility accruing to the agents in the model, it, however, does influence the decisions undertaken.

**The Principle of Credible Promises**

The crucial line of analysis in the model is the presence of what the authors have termed *credible commitment* on the part of the two candidates so considered (Aragonès et al., 2007). By that they mean, in equilibrium the voters believe only those campaign messages which are within the maximal range of incentive compatible promises for both the candidates. That is, only those promises will be believed in equilibrium which the candidates have an incentive to fulfill once elected to a term in office. In the presence of the discount factor $\delta$ for a particular candidate, the model predicts a *d($\delta$)* which the voters believe, if and only if the difference in the position of the candidate's ideal point and his promise is less than *d($\delta$)* (Aragonès et al., 2007).

Thus, in the equilibrium described earlier, voters will believe the entire set of promises coming from a candidate if the distance between his ideal point and his proposed policy is not more than *d($\delta$)*. However, if a promise falls outside this range of *credible promises* or the candidate is one with a *bad* reputation (that is, he had reneged on his promised platform at least once before) or he makes no promises at all, the voters will believe that once elected, the candidate will end up implementing his ideal point. That is, the absence of an *incentive-compatible* promise induces the voters to predict that once in office the candidate will choose his bliss (or ideal) point as his preferred policy.

At the office stage of the election, the winning candidate will implement his policy that maximizes his expected payoffs. This is done after considering the voters' strategies for all future elections, which in turn depends upon the candidate's promised policy platform and his post-election policy choice. Hence, candidates at this stage compare the costs and benefits of reneging. The cost from reneging pertains to the difference in the expected future payoffs of a candidate with a *good* reputation and that of a candidate with a *bad* one, while, the gains from reneging are reflected by the immediate increase in the utility of the candidate by deviating from his promise and choosing his ideal point (Aragonès et al., 2007). Thus, a candidate will renege only if the benefit of reneging (given by the instantaneous rise in the utility) is more than his expected future loss in payoffs (given by the fact that he would now be believed as a candidate with a *bad* reputation for all future elections to come) (Aragonès et al., 2007).

In the ensuing equilibrium, candidates come up with those policies that are incentive com-patible with them and duly follow up on them, if elected. Therefore, voters knowing this will believe the promises being made and will vote for the candidate whose policy promise lies closest to the median voter's ideal point. Thus, the winning strategy is one where a candidate's promised policy is at least as appealing to the median voter as is his opponent's policy.

The *candidates'* strategies can thus be written down as -

(a). If neither candidate has ever reneged on a promise, then the candidate whose ideal point is farther from the median voter's ideal point promises the policy that is closest to the median voter's ideal point consistent with incentive-compatibility. The candidate whose ideal point is closer to the median voter's ideal point promises a policy that is equally attractive to the median voter. If elected, both candidates fulfill their promise.

(b). If both candidates have reneged on a promise in the past, then both candidates promise to implement the median voter's ideal point. If elected, they implement their own ideal point.

(c). If one candidate has reneged on a promise but the other candidate has never reneged, the candidate who has reneged promises to implement the median voter's ideal point. If elected, he implements his own ideal point. The candidate who has not reneged promises a policy that is as attractive to the median voter as the opponent's ideal point, if such a promise is incentive-compatible. If that promise is not incentive-compatible, then he promises his ideal point. If elected, he fulfills his promise.

*Voters' Strategies* Each voter votes for the candidate whose promised platform, if elected, would help maximize their own future expected utility. The beliefs of the voters' can then be written down as -

(a). Voters believe incentive-compatible promises of *good* candidates will always be fulfilled.

(b). Voters believe that a promise which is not incentive-compatible will never be fulfilled.

(c). Voters believe that a *bad* candidate will always implement his ideal point if elected, irrespective of his pre-election promised policy.

These set of strategies constitutes a sub-game perfect equilibrium.

PROPOSITION 1. *The strategies just described constitute an equilibrium. The promises believed and fulfilled in equilibrium with linear utility functions are those within d\*(δ) of the candidates' ideal points, here -*[4]

$$d^*(\delta) = \begin{cases} 0 \text{ if } 0 \leq \delta \leq \frac{1}{2} \\ \frac{3}{2}(1 - \sqrt{\frac{4-5\delta}{3\delta}}) \text{ if } \frac{1}{2} \leq \delta \leq \frac{3}{4} \\ 1 \text{ if } \frac{3}{4} \leq \delta \leq 1 \end{cases} \qquad (2)$$

In equation (2), the distance $d^*(\delta)$ characterizes the maximal range of incentive compatible promises in equilibrium. Thus, one can observe that in equilibrium voters believe all the promises of a *good* candidate (one who has never reneged), who in turn themselves fulfill all the promises they had made. Hence, there exists a continuum of equilibria, such that $\forall d \leq d^*(\delta)$ there remains a equilibrium in which all the voters believe a promise up to a distance $d$ from the candidate's bliss point (Aragonès et al., 2007).

In tune to the same, an interesting comparative statics result that comes out from the benchmark model is that -

$$\frac{\partial d^*(\delta)}{\partial \delta} = \frac{1}{\delta^2}\sqrt{\frac{3\delta}{4-5\delta}} \geq 0 \qquad (3)$$

The above equation shows us that the maximal incentive-compatible promises believed in equilibrium is an increasing function of the discount factor. That is, as the discount factor increases, the value of credibility on the part of the candidates' also increases (Aragonès et al., 2007). This implies that larger promises will be both kept by candidates as well as believed by voters, in equilibrium. A point to note in this regards is that a candidate's expected flow of payoffs - given by $v_G$ - of keeping up with a *good* reputation is independent of whether his rival is *good* or *bad* (Aragonès et al., 2007). This is shown in the form of the

---

[4]For an elaborate discussion on the mathematical derivation of the results, please see Appendix A.1. of Aragonès et al. (2007)

equation -

$$v_{GG}(d^*(\delta)) - v_{BG}(d^*(\delta)) = v_{GB}(d^*(\delta)) - v_{BB}(d^*(\delta)) \qquad (4)$$

where the two subscripts (beginning from left to right) denotes - whether a candidate is *good* or *bad* concerning the reputation of his rival respectively; in equilibrium this arises because of the assumption of linearity of the utility functions.

Thus, from equation (13) we can further conclude that -

$$C^D(d;\delta) - C^S(d^S;\delta) = d^S(\delta)[(\text{Aragonès et al., 2007})]$$

That is, the cost of reneging when both the candidates are *good* equals the cost of reneging when one of the candidate is *good* and the other is *bad*. This necessarily implies that $d^D(\delta)$ = $d^S(\delta)$.[5] That is, the maximal promises believed by voters and fulfilled in by candidates (in equilibrium) are the same for the winning candidate irrespective of whether his opponent is *good* or *bad*.

## Non-naive Voting Behavior

### Overview

The primary assumption made in the benchmark model of Aragonès et al. (2007) is the presence of myopic voting behavior on the part of the voters. Under the premise of complete information with infinitely-repeated sequential elections, the result that the model duly shows us is the one where the range of maximal credible promises- also the subgame perfect equilibrium of the game- given by $d^*(\delta)$ is equal among both the cases, where (a). Both the candidates carry a *good* reputation, and (b). One of them carries a *good* with the other a *bad* reputation. Mathematically, which implies that - $d^D(\delta) = d^S(\delta)$ (Aragonès et al., 2007).

However, the question that then arises is - what would happen if the assumption of myopic voting behavior is supplanted by one that of strategic behavior? That is, how would the maximal range of credible promises change if the voters much like the candidates start not only to discount their future payoffs, but also induce the latter to make even better, more credible promises in equilibrium than before. In the benchmark model, it was shown that the moderate (winning) candidate had only to offer the representative voter a policy platform that was at least as attractive as the one proposed by his rival (the extreme candidate).

---

[5] See Appendix A.1. of Aragonès et al. (2007) for a more detailed explanation of the same.

Under the setup put forth in the benchmark model, voters do not discount the future. They in turn form the basis of a candidate's reputation as one of *good* as long as $x = p$, where 'x' stands for the implemented policy and 'p' the promise. Anticipating correctly the post-election behavior of candidates (both *good* and *bad*), they elect the one whose policy lies closer to their own ideal point. That is, for a *bad* candidate they correctly anticipate that the implemented policy would be the candidate's own ideal point, while for a *good* candidate, the policy implemented would match his promise as long as $|x_i - p| \leq d^*(\delta)$, where $x_i$ is the candidate's bliss point. In my present work, instead of treating voters as one-shot utility maximizers, I regard them to be more *strategic* than their counterparts in the benchmark model. Since voters would believe all the promises made by the *good* candidate within a distance d* of the ideal points [that is, $p \leq x_i - d^*(\delta)$ for a right-leaning candidate and $p \geq -x_i + d^*(\delta)$ for a left-leaning one], voters are willing to push the candidates to the upper bound of what they can promise without being incentive incompatible. Doing so not only maintains the reputation of the candidate concerned but more importantly, unambiguously makes the voters better-off than before - a Pareto improvement. Given such an assignment, a candidate has no option but to promise his highest possible incentive-compatible policy to the voters, since failure to do so would result in a permanent *bad* reputation. Thus, a candidate (in order to maintain his *good* reputation) calls out his maximal incentive-compatible promise - which he would end up implementing, if elected to office.

Therefore, the presence of *non - naive voting behavior* would mean that the moderate (winning) candidate would now be induced to promise his maximum possible incentive-compatible platform or bear the loss of credibility in all future elections to come. Thus, some additional equilibria are possible, if the voters not only start to discount their future payoffs but also begin to behave more strategically in ways described earlier.

## An Example of Strategic, Non-naive Voting Behavior

An example of how non - naive voting behavior might lead to some additional equilibria can be shown as follows -

Let us consider the calculation of the term $v_{GG}$[6], which in the benchmark model has been shown to be equal to -

$$v_{GG}(d) = \int_0^{1-d} \int_{-1}^{-x_R-d} u_L(x_R)\, dx_L\, dx_R + \int_d^1 \int_{-x_R+d}^0 u_L(x_L)\, dx_L\, dx_R + \int_0^d \int_{-d}^0 u_L(0)\, dx_L\, dx_R$$

$$+ \int_d^1 \int_{-x_R}^{-x_R+d} u_L(-x_R + d)\, dx_L\, dx_R + \int_{-1}^{-d} \int_{-x_L-d}^{-x_L} u_L(-x_L - d)\, dx_R\, dx_L = -\frac{1}{2} \quad (5)$$

Reconsidering the fourth term from the left of the equation, we find that under the premises of myopic voting behavior, if the ideal point of the 'Left' candidate (also the moderate one in this case) lies between the ideal point of candidate 'R' (i.e., $x_R$)[7] and that of the maximum (credible) platform '$(-x_R + d)$'; then the moderate (*good*) candidate always wins the election by promising a platform equal to '$(-x_R + d)$', which is at least as attractive to the median voter as the one promised by candidate 'R'.

However, with non - naive voters (Aragonès et al., 2007) the moderate candidate cannot make do by just proposing a policy platform equal to '$(-x_R + d)$'. Now they ask the moderate candidate that, "Since you can credibly promise a platform to the extent of '$(x_L + d)$'[which is closer than '$(-x_R + d)$' is to the ideal point of the median voter], why don't you do that and go the fullest extent possible (credibly) and promise us '$(x_L + d)$' instead of just '$(-x_R + d)$'? If this is indeed the case, then you better promise us '$(x_L + d)$', otherwise, not only shall we not vote for you in this election but also make sure your reputation (in our

---

[6]Denotes the one-shot election expected utility for a candidate when both of them have a *good* reputation.
[7]-'$x_R$'is the mirror image of the point '$x_R$'on the positive spectrum of the real line.

eyes) turns *bad* for all future elections to come. The choice is now yours to make."

The moderate candidate hearing this is now in a dilemma - should he stay put with his bid of fulfilling '(-$x_R$ + d)'or pay heed to the voters' words of caution by promising to fulfill '($x_L$ + d)'. If he does pay heed, then he will both win the election as well as maintain his reputation as a *good* candidate for all future elections to come, but at the same time would suffer a loss in utility from moving further away from his ideal point. If the candidate decides to maintain his reputation of staying *good* (paying heed to the non - naive voters), then the integral in question becomes[8] -

$$\int_{d}^{1} \int_{-x_R}^{-x_R+d} u_L(x_L + d) \, dx_L \, dx_R \qquad (6)$$

The change to note in the above integral [with respect to the one in equation (7)]is that now, the moderate (winning) candidate's promised policy platform is equal to his maximum credible promise of '($x_L$ + d)'(as induced by the non - naive voters) instead of '(-$x_R$ + d)'(when the voters where naive as in the benchmark analysis). In the same vein, the above line of reasoning can be applied to each of the other individual integrals of $v_{GG}$ as well as to those in $v_{GB}$, $v_{BB}$ and $v_{BG}$ - whenever the situation arises of the voters having the possibility to induce the moderate candidate in promising his maximum credible platform, even when the more extreme (losing) candidate keeps on promising his highest credible platform as shown in the benchmark model.

The strategies for the equilibrium can be described as follows -

*Candidates'* strategies :

1. If neither candidate has ever reneged on a promise, then the candidate whose ideal point is farther from the median voter's ideal point promises the policy that is closest to the median voter's ideal point consistent with incentive compatibility. The candidate whose ideal point is closer to the median voter's ideal point promises a policy that is equally attractive to the median voter. If elected, both candidates fulfill their promise (Aragonès et al., 2007).

---
[8]This will hold only if -
$$(-x_R + d) + d \leq 0 \implies x_R \geq 2d$$

2. If both candidates have reneged on a promise in the past, then both candidates promise to implement the median voter's ideal point. If elected, they implement their own ideal point.

3. If one candidate has reneged on a promise but the other candidate has never reneged, the candidate who has reneged promises to implement the median voter's ideal point. If elected, he implements his own ideal point. The candidate who has not reneged promises a policy that is as attractive to the median voter as the opponent's ideal point, if such a promise is incentive compatible. If that policy is not incentive-compatible, then he promises his ideal point. If elected, he fulfills his promise.

4. If a candidate has never reneged on a promise in the past, and his implemented policy (also his promise to the voters) is at least as much close to that of the median voter as of his opponent but less than the maximal credible distance ($d^*$) he can move. Then, under non - naive voting behavior, he will implement the policy that lies at a point equal to the maximal credible distance away from his ideal point. He does so in order to maintain his reputation as a *good* candidate.

*Voters'* strategies - Each voter casts his or her vote for the candidate whose expected policy, if elected, maximises the voter's utility. The strategies are as follows:

1. Voters believe that incentive-compatible promises of candidates who have never reneged on a promise will be fulfilled.

2. Voters believe that a candidate who makes a promise that is not incentive compatible will implement his ideal point.

3. Voters believe that a candidate who has reneged on a promise in the past will implement his ideal point.

4. Voters enforce their common belief that if a candidate can move to a point equal to the maximal credible distance away from his ideal point and win the election, then the candidate must implement the above policy or he risks losing his credibility for the rest of all future elections.

PROPOSITION 2 *The strategies just described constitute an equilibrium. The promises believed and fulfilled in equilibrium with linear utility functions are those within d\*(δ) of the candidates' ideal points, where,*

$$d_B^*(\delta) = \begin{cases} 6(1 - (1/\delta)) \text{ if } 0 < \delta < 1 \\ \\ 0 \text{ if } \delta = 1 \end{cases} \quad (7)$$

and,

$$d_G^*(\delta) = \begin{cases} 0 \text{ if } 0 \leq \delta \leq \frac{3}{5} \\ \\ \sqrt{3}\sqrt{(3 - \frac{2}{\delta})} \text{ if } \frac{3}{5} \leq \delta \leq 1 \end{cases} \quad (8)$$

Each of the $d^*$s characterizes the maximal range of credible promises in equilibrium. By credible or incentive compatible promises, we mean that there exists an equilibrium where the voters believe promises made by the candidates up until a distance of $d$ from the latter's respective ideal points.

The analysis can also be extended to find out the comparative statics results for each of $d_G^*$ and $d_B^*$. We find that

$$\frac{\partial d_G^*}{\partial \delta} = \sqrt{\frac{3\delta}{3\delta - 2}} \cdot (\frac{1}{\delta^2}) > 0 \quad (9)$$

and,

$$\frac{\partial d_B^*}{\partial \delta} = (\frac{1}{\delta^2}) > 0 \quad (10)$$

Thus, we observe that the maximal promise believed in equilibrium for both the above scenarios is an increasing function of the discount factor, $\delta$. That is, the worth of maintaining the reputation increases with increase in the discount factor, which in turn implies that larger promises will be both believed and implemented in equilibrium. This result also leads us to an interesting aspect of this model - in the sense that - candidates who are more probable of winning the election for that particular term are more likely to keep their promises, and in turn the voters are more likely to believe such promises in equilibrium. As Aragonès et al. (2007) puts it succintly, ".. promises are more likely to be believed at the same time that candidates are more likely to make them."

Interestingly, we note that even though the value of $d_G^*$ lies less than that of Aragonès et al. (2007)'s result - for higher values of the discount factor the two converges to the same value of 1. Although, we must note here that $d_B^*$ always lie below both $d_G^*$ and $d_A^*$ (Aragonès et al., 2007).

## Mechanics underlying the Results

I shall begin this section by describing each of the sub-integrals of both $V_{BB}$ vs $V_{GB}$ and $V_{BG}$ vs $V_{GG}$. While elaborating on them, I shall also be pointing out the specific reason as to how and why the notion of non - naive voting behavior is incorporated in each of these integrals at hand. After which, I will present a graphical representation of each of the results of equation (3) and (4) respectively, where, I will show the variation in the curvature of each of the respective graphs for three (3) different sets of values of $\delta$. Finally, I will present a combined graphical representation of all the three results that we have viz. that of Aragonès et al. (2007), $d_G^*$ from $V_{BG}$ vs $V_{GG}$, and $d_B^*$ from $V_{BB}$ vs $V_{GB}$, respectively, with its resultant implications.

1. $V_{BB}$ vs $V_{GB}$

In case of the calculation of $v_{BB}$, as shown in the benchmark model, the one-shot utility accruing to the *bad* candidate when his opponent is also of *bad* reputation has been shown to be -

$$v_{BB}(d) = \int_0^1 \int_{-1}^{-x_R} u_L(x_R)\,dx_L\,dx_R + \int_0^1 \int_{-x_R}^{0} u_L(x_L)\,dx_L\,dx_R$$

However, applying the logic used in the above example, we can easily see that the assumption of non - naive voting behavior would not lead to any changes in the integrals as shown in the above equation. This result is due to the fact that for a candidate with a bad reputation, the voters (whether strategic or otherwise) would always believe that whatever be the promised policy of the candidate, in practice he would always end up implementing his own ideal point. Hence, the question of coming up with credible promises in this scenario doesn't arise at all. That is, the one-shot utility in case of $v_{BB}$ remains the same under both strategic and naive voting behavior.

Alternatively, in case of $v_{GB}$ (the one-shot election utility accruing to the candidate with a good reputation when his rival has a bad reputation), according to the benchmark model we have -

$$v_{GB}(d) = \int_0^{1-d} \int_{-1}^{-x_R-d} u_L(x_R)\,dx_L\,dx_R + \int_0^{1-d} \int_{-x_R-d}^{-x_R} u_L(x_L)\,dx_L\,dx_R + \int_{1-d}^{1} \int_{-1}^{-x_R} u_L(-x_R)\,dx_L\,dx_R$$

$$+ \int_0^1 \int_{-x_R}^{0} u_L(x_L)\,dx_L\,dx_R = -\frac{1}{6} - \frac{(1-d)^3}{3} \tag{11}$$

Now, for each of these integrals we will proceed by describing the underlying mechanism at work - first, under a naive voting scenario (Aragonès et al., 2007), and then under the assumption of non - naive voting behavior on part of the voters. They are as follows -

(a). $x_R \in [0, (1-d)]$, $x_L \in [-1, (-x_R- d)]$ :

As shown by Aragonès et al. (2007), candidate L always loses irrespective of the position of his rival's ideal point. This result is because L's ideal point lies to the left of the least (effective) possible bliss point of his rival candidate, that is '$-x_R - d$'. So, even if L promises a policy '$x_L + d$' (his maximum credible promise), he won't be able to overcome the platform promised by R (which in this case is his own ideal point). So, under non - naive voting behavior too, L will end up losing the election because even if the voters induce L to promise his maximum credible platform (equal to '$x_L + d$') he would still end up short of R's ideal point.

(b). $x_R \in [0, (1\text{-}d)]$ , $x_L \in [-x_R - d, -x_R]$ :

This is the exact reverse situation of the case we have discussed above. Here, under naive voting behavior, candidate L can win the election by promising a platform equal to '$x_R$', which is at least as close to the median voter as the one promised by R (who has a bad reputation). However, under non - naive voting behavior, candidate L can be induced to promise a platform '$x_L + d$' (which is his maximum credible promise given his ideal point). If L indeed puts forth his maximum credible policy platform (in equilibrium), he will win the election. This will happen because L's promised policy would always be more than the ideal point of R. The integral will then become -

$$\int_0^{1-d} \int_{-x_R-d}^{-x_R} u_L(x_L + d) \, dx_L \, dx_R$$

(c). $x_R \in [1\text{-}d, 1]$ , $x_L \in [-1, -x_R]$ :

As shown in equation (18), candidate L wins the election (under the setting of naive voting behavior) by promising a platform equal to '-$x_R$', which is at least as good as the platform put forth by candidate R (who is the one with the bad reputation). However, the presence of non - naive voting behavior would mean that now the voters would in turn induce the *(good)* candidate L in promising a platform equal to '$-x_L + d$' instead of just staying put with '$-x_R$'. Under such a circumstance, the integral in question will then become -

$$\int_{1-d}^{1} \int_{-1}^{-x_R} u_L(x_L + d) \, dx_L \, dx_R$$

(d). $x_R \in [0, 1]$, $x_L \in [-x_R, 0]$ :

As shown in Aragonès et al. (2007), candidate L will always win the election (by implementing '$x_L$') irrespective of where R's ideal point lies. This result can be seen from the range of each candidate's positioning of ideal points as in equation (11). Since candidate L's ideal point will always lie to the right (and thus closer to the median voter's ideal point), he can win the election by promising a platform equal to his own bliss point. The voters', in turn, would vote for L, believing in his promise of implementing his ideal point. However, under a non - naive voting environment, the candidate (here L) would be induced to promise a platform equal to '$(x_L + d)$' instead of just '$-x_R$' [9]. The integral will then become -

$$\int_0^1 \int_{-x_R}^0 u_L(x_L + d)\, dx_L\, dx_R$$

Thus, under non - naive voting behavior the *new* integrals can be collectively be written down as -

$$v_{GB}(d) = \int_0^{1-d} \int_{-1}^{-x_R-d} u_L(x_R)\, dx_L\, dx_R + \int_0^{1-d} \int_{-x_R-d}^{-x_R} u_L(x_L+d)\, dx_L\, dx_R + \int_{1-d}^1 \int_{-1}^{-x_R} u_L(x_L+d)\, dx_L\, dx_R$$

$$+ \int_0^1 \int_{-x_R}^0 u_L(x_L + d)\, dx_L\, dx_R = -\frac{d^2}{6} - \frac{1}{2} \quad (12)$$

Then, the expected future payoffs for a *good* candidate given his opponent being *bad* is equal to :

$$V_{GB}(d; \delta) = \sum_{t=1}^\infty \delta^t v_{GB}(d)^{10}$$

And similarly, the expected future payoffs for a *bad* candidate given his opponent being also

---
[9] Applying the same logical premise used to derive the integral in equation (1)
[10] Aragonès et al. (2007)

of *bad* reputation as well, is given by -

$$V_{BB}(d; \delta) = \sum_{t=1}^{\infty} \delta^t v_{BB}(d)^{11}$$

Then the cost of reneging given by $C^S(d; \delta)$ is a function of the maximal credible promise believed by voters and the discount factor. Such that it is equal to -

$$C^S(d; \delta) = V_{GB}(d; \delta) - V_{BB}(d; \delta) = \frac{\delta}{1-\delta} \cdot \frac{-d^2}{6} \quad (13)$$

2. $V_{GG}$ vs $V_{BG}$

In case of the calculation of $v_{GG}$, as shown in the benchmark model, the one-shot utility accruing to the *good* candidate when his opponent is also of *good* reputation has been shown to be -

$$v_{GG}(d) = \int_0^{1-d} \int_{-1}^{-x_R-d} u_L(x_R)\, dx_L\, dx_R + \int_d^1 \int_{-x_R+d}^0 u_L(x_L)\, dx_L\, dx_R + \int_0^d \int_{-d}^0 u_L(0)\, dx_L\, dx_R$$

$$+ \int_d^1 \int_{-x_R}^{-x_R+d} u_L(-x_R + d)\, dx_L\, dx_R + \int_{-1}^{-d} \int_{-x_L-d}^{-x_L} u_L(-x_R - d)\, dx_R\, dx_L = -\frac{1}{2} \quad (14)$$

Now, for each of these integrals we will proceed by describing the different scenarios under the assumption of non - naive voting behavior on part of the voters. They are as follows -

(a). $x_R \in [0, 1\text{-}d]$, $x_L \in [-1, -x_R - d]$ :

Applying the premises of non - naive voting behavior as applied in the earlier scenarios, we find that the implemented policy in this case remains same as that of Aragonès et al. (2007). Thus, the integral remains the same as in equation (21).

---
[11] Aragonès et al. (2007)

(b). $x_R \in [d, 1]$, $x_L \in [-x_R + d, 0]$:

In this scenario, with the given range of integrals in question, we find that voters (strategic) will force the candidate to both promise, and hence implement his maximal possible credible policy of '$x_L + d$. The *new* integral then becomes -

$$\int_d^1 \int_{-x_R+d}^0 u_L(x_L + d)\, dx_L\, dx_R$$

(c). $x_R \in [0, d]$, $x_L \in [-d, 0]$:

In this sub-integral, we find that the implemented policy in this case remains same as that of Aragonès et al. (2007), even under non - naive voting behavior.

(d). $x_R \in [d, 1]$, $x_L \in [-x_R, -x_R + d]$:

This part has already been described under the section titled 'An Example of non - naive voting', and the new integral culminating from the logical premise described in the aforementioned section has been shown in equation (2) above.

(e). $x_R \in [-1, -d]$, $x_L \in [-x_L - d, -x_L]$:

In this case, with the given range of integrals in question, we find that the presence of non - naive voting will make the *good* candidate to both promise, and hence implement '$x_R - d$'. The *new* integral then becomes -

$$\int_{-1}^{-d} \int_{-x_L-d}^0 u_L(x_R - d)\, dx_R\, dx_L$$

Thus, under non - naive voting behavior the *new* integrals can be collectively be written down as -

$$v_{GG}(d) = \int_0^{1-d}\int_{-1}^{-x_R-d} u_L(x_R)\,dx_L\,dx_R + \int_d^1\int_{-x_R+d}^{0} u_L(x_L+d)\,dx_L\,dx_R + \int_0^d\int_{-d}^{0} u_L(0)\,dx_L\,dx_R$$

$$+ \int_d^1\int_{-x_R}^{-x_R+d} u_L(x_L+d)\,dx_L\,dx_R + \int_{-1}^{-d}\int_{-x_L-d}^{-x_L} u_L(x_R-d)\,dx_R\,dx_L = \frac{(1-d)^3}{2} - \frac{d^2}{2} + d - 1 \quad (15)$$

Alternatively, in the case of $v_{BG}$, we find that upon applying of logical premises of non-naive voting behavior, the sub-integrals remains the same as that of Aragonès et al. (2007). The integral can then be written down as -

$$v_{BG}(d) = \int_0^1\int_{-1}^{-x_R} u_L(x_R)\,dx_L\,dx_R + \int_d^1\int_{-x_R}^{-x_R+d} u_L(-x_L)\,dx_L\,dx_R + \int_0^d\int_{-x_R}^{0} u_L(-x_L)\,dx_L\,dx_R$$

$$+ \int_d^1\int_{-x_R+d}^{0} u_L(x_L)\,dx_L\,dx_R = -\frac{5}{6} + \frac{(1-d)^3}{3} \quad (16)$$

Then, the expected future payoffs for a *good* candidate given his opponent being *good* is equal to :

$$V_{GG}(d;\delta) = \sum_{t=1}^{\infty} \delta^t v_{GG}(d)$$

And similarly, the expected future payoffs for a *bad* candidate given his opponent being also of *bad* reputation as well, is given by -

$$V_{BG}(d;\delta) = \sum_{t=1}^{\infty} \delta^t v_{BG}(d)$$

Then the cost of reneging given by $C^S(d;\delta)$ is a function of the maximal credible promise believed by voters and the discount factor. Such that it is equal to -

$$C^S(d;\delta) = V_{GG}(d;\delta) - V_{BG}(d;\delta) = \left(\frac{\delta}{1-\delta}\right)\left(\frac{(1-d)^3}{6} - \frac{d^2}{2} + d - \frac{1}{6}\right) \tag{17}$$

Now, with the respective *Costs of Reneging* available to us from equations (20) and (24) respectively, we are able to derive the respective $d^*$s each for both the cases, as given by the equations (3) and (4) respectively. This is done by equating the Cost of Reneging to the Gain from Reneging as has been shown earlier.

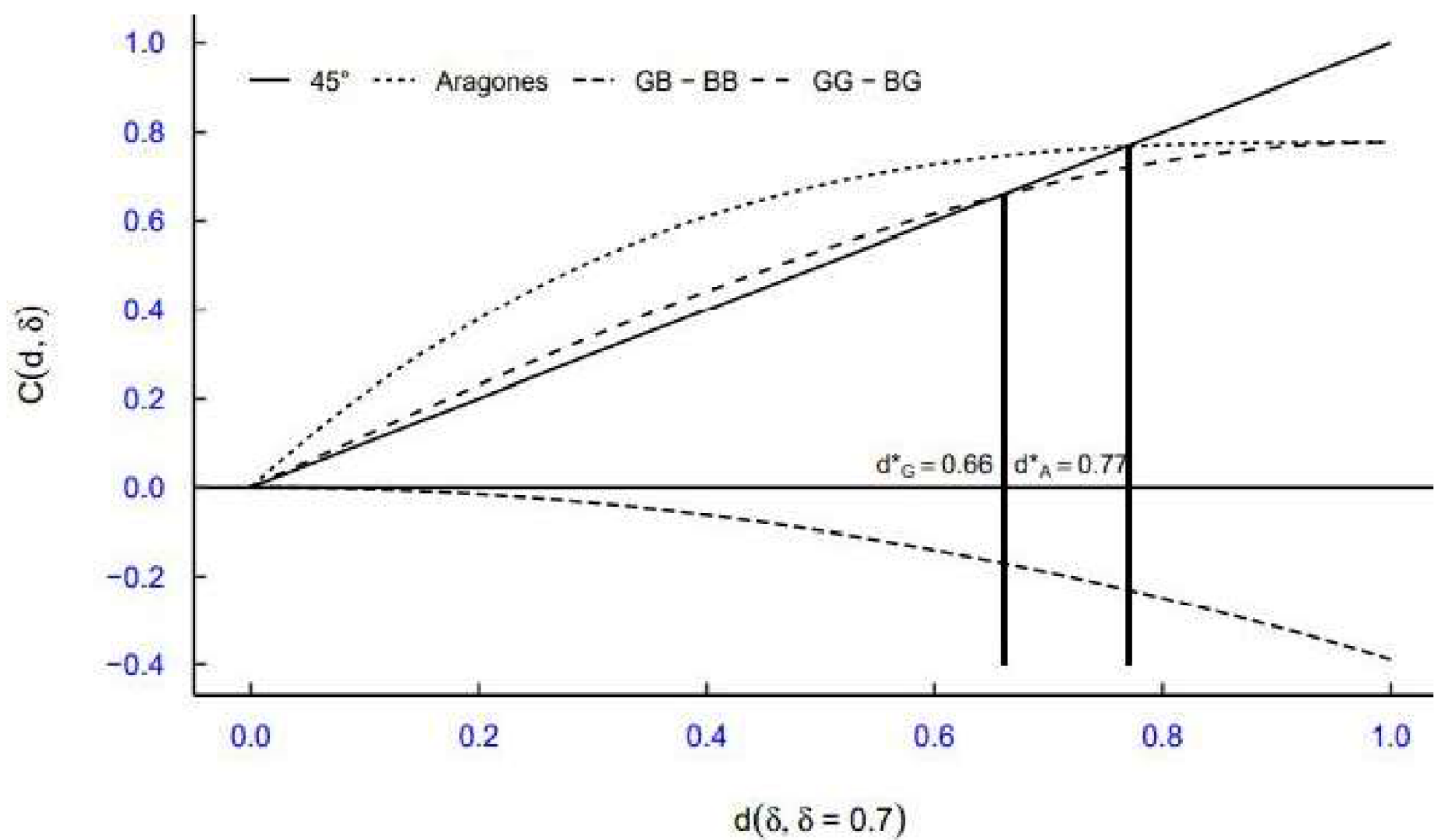

Figure 1: Comparative analysis of $d_A^*$, $d_G^*$, and $d_B^*$ for $\delta = 0.7$

As can be seen from Figure 2.1 below, the respective values of the maximal incentive-compatible promises (for $\delta = 0.7$) are ordered in the following ascending order of values-

$$d_B^* < d_G^* < d_A^*$$

where, $d^*_A$, $d^*_G$, and $d^*_B$ each denotes the equilibrium maximal credible promise for Aragonès et al. (2007) (as shown by point $d^*_A$ in the figure), when a candidate's opponent is *good* (shown by point $d^*_G$ in the figure), and when the opponent is assumed to *bad*. We must note here that with increase in the value of the discount factor the value of $d^*_G$ converges to the Aragonès et al. (2007)'s result. In fact for the range of $\delta \epsilon [0.75, 1]$ - both the values are same and equals 1.

## The Intuition

A voter in the benchmark model of Aragonès et al. (2007) is assumed to be *naive*, in the sense that he not only disregards the future while calculating his utility in each period but also chooses the candidate who offers him his preferred platform (Aragonès et al., 2007). This premise takes place only because there is no link between what platform is promised and/or enacted except for the extent of its enactment (Aragonès et al., 2007).

However, under non - naive voting behavior, the same voter now, not only includes the future into his utility analysis but also induces the more moderate candidate to promise a better (or more credible) promise or face the risk of suffering a *bad* reputation for the rest of the game's history. Such behavior on part of the individual voter would now induce the moderate candidate to be more risk-averse and would lead to a reduction in both the value of maintaining a reputation as well as the maximal incentive compatible (credible) promise he would end up making in equilibrium.

## Implication of the Results

As we can see from both the intuition and the results of incorporating *non - naive voting behavior* instead of *naive voting* - the voters are *better* able to bargain for not only a much more credible promise from the candidate in question but more importantly is able to reduce the chances of cheap-talk political campaigns, where rhetoric often trumps reality. This latter aspect of political behavior on part of candidates often highlights the vagueness and ambivalence in our daily political life (Condor et al., 2013). Including *strategic* behavior thus shows us that an electoral mechanism of such nature can reduce such vague arguments by creating a credible system of accountability of what candidates *say* and *do* if they are elected to power. However, the above implication carries a caveat in the form that the result is a reflection of a crucial assumptions of complete information and preference uniformity of the agents in our present analysis. The simple fact that the presence of asymmetry in both information and preference marks our daily political behavior *might* not reflect the same implications of what our present study has achieved to show so far. Although they do suggest that such asymmetries are all but relevant fuel for performing further research on this agenda.

Electoral competition provides the backbone to the functioning of strong legislative institutions in any democratic country. Yet the dynamic nature of such competition provides an unique opportunity to a scholar to untangle the challenges posed by the interaction between the candidates and the voters. So far, such multifarious forms of interactions have been ably captured via game-theoretic models, that have sought to understand voting behavior under the context of repeated elections, electoral accountability, punishment strategies, incumbency (dis)advantages, et cetera. In turn, the present analysis attempted to extend the benchmark model of Aragonès et al. (2007) on one singular count -

1. Ceteris paribus, by including non - naive voting behavior in place of naive (or myopic) voting structure of the benchmark model.

The results so derived above have supported the intuitive premises made (but not solved) by Aragonès et al. (2007) in their benchmark analysis. We find that under non - naive voting behavior, where a voter induces the more moderate candidate to promise and implement a better (or more credible) campaign platform (or suffer loss of credibility) - the maximal credible promise decreases because the value of the reputation to the candidate falls along with it.

Interestingly, future studies could look at the above behavior of voters from the same equilibrium-selection criterion aspect of analysis. That is, instead of viewing voter behavior from a 'myopic-strategic' angle, it might helpful to see it from an equilibrium-selection criterion - whereby out-of-equilibrium restrictions can be imposed to deter candidates from announcing policies that fails to satisfy the incentive compatibility condition of the voters. Such a work can be done in a setting of incomplete information pervading the electoral arena, where the politicians' policy preferences are private information, and voters are able to gauge reputations via *signals* from candidates in the form of electoral promises and past performances. In line with the equilibrium-selection view of analysis, we can also think of other beliefs of non - naive voters, where the function regarding expectation fulfilment follows some function that is inversely proportional to the distance of the announced policy and the ideal point of the candidate. In tune, a complete characterization of the nature of equilibria in the generalized setup might be of interest, which may generate the specific equilibrium obtained in this work as a limiting case.

# Limited Punishment

## Overview

It was assumed in the benchmark model of Aragonès et al. (2007) that once a candidate reneges upon his promised policy platform, voters would continue to punish him for the rest of all future elections by not believing in any of his future promises. There exists other equilibria if this assumption is supplanted by that of limited punishment, where, after a candidate reneges once, voters punish him for only a finite number of periods after which his promises are again believed. The intuition behind employing a limited punishment analysis is to see whether the alteration in the basic assumption of the infinite punishment strategy for having a *bad* reputation duly affects the incentive compatible policy announcements from the two candidates. It makes sense that alteration in the punishment strategy would lower the cost of reneging (as reneging would be punished only for a finite period of time by the *non - naive voters*). Keeping the above concept in our midst, I will show that the equilibrium level of maximal credible promises ($d^*$) under limited punishment (each for k = 1, 2 and 3) is *lower* than that of the result found by (Aragonès et al., 2007).

The strategies for the equilibrium described are as follows :

*Candidates'* strategies -

1. If neither candidate has ever reneged on a promise, then the candidate whose ideal point is farther from the median voter's ideal point promises the policy that is closest to the median voter's ideal point consistent with incentive compatibility. The candidate whose ideal point is closer to the median voter's ideal point promises a policy that is equally attractive to the median voter. If elected, both candidates fulfill their promise.

2. If both candidates have reneged on a promise in the past, then both candidates promise to implement the median voter's ideal point. If elected, they implement their own ideal point.

3. If one candidate has reneged on a promise but the other candidate has never reneged, the candidate who has reneged promises to implement the median voter's ideal point. If elected, he implements his own ideal point. The candidate who has not reneged promises a policy that is as attractive to the median voter as the opponent's ideal point, if such a promise is incentive compatible. If that policy is not incentive-compatible, then he promises his ideal point. If elected, he fulfills his promise.

*Voters' strategies* - Each voter casts his or her vote for the candidate whose expected policy, if elected, maximises the voter's utility. The strategies are as follows:

1. Voters believe that incentive-compatible promises of candidates who have never reneged on a promise will be fulfilled.

2. Voters believe that a candidate who makes a promise that is not incentive compatible will implement his ideal point.

3. Voters believe that a candidate who has reneged in the past will always implement his or her ideal point. This belief is held over the period of punishment after which the candidate's reputation is restored back to one of good.

With the respective strategies in place, we will now proceed by first elaborating upon the underlying workings for k=1, followed by its results (and the economic intuition behind it), and also those for k equalling 2 and 3 respectively.

## Results for k=1

We begin by assuming that the period of punishment imposed by the voters is equal to one (i.e. k=1). This means that if a candidate reneges on a promise today (let us assume it to be time 't'), the voters will punish him (by not believing his promises) for the time period starting from 't+1' till 't+2', after which (i.e. from time ('t+3')) his reputation is restored to one of *good*. We then proceed to calculate the cost of reneging (i.e. the discounted stream of future expected pay-offs if the candidate decides to renege)[12], and therefore the equilibrium $d^*$ for a *good* candidate. This idea is worked out once - by assuming that his opponent is *good* - and then by assuming that his opponent is *bad*. Interestingly, we find that the equilibrium condition[13] for k=1, and later for k equals to 2 and 3, respectively, to be converging to the equilibrium condition of Aragonès et al. (2007) as k tends to infinity.

| Time      | 0 | 1     | 2     | 3 | 4 |
|-----------|---|-------|-------|---|---|
| Candidate | G | G (B) | G (B) | G | G |
| Opponent  | G | G     | G     | G | G |

Using the methodology propounded by Aragonès et al. (2007)- where the *cost of reneging* is calculated each for opponent being *good* (and when he or she is *bad*.) and is equalled to the *gain from reneging* (given by d) each for the two cases respectively. As seen in the table above, the first row shows us a slice of the infinitely-repeated game where t =0 being the period of time where the elected candidate is deciding whether to renege or not. We see that at t=0, both the candidates' are *good*, and the winning candidate is deciding whether to renege (and turn *bad*) or to not renege, given his opponent being *good*. If he decides to renege, then at t=1, his reputation is shown to be *bad* (as given by B in the brackets of the above table). Now, since the punishment period is one, he stays *bad* for period 1 to 2, after which his reputation is restored to *good* by the voters.

Mathematically, using the one step deviation principle, the above decision can be shown to

---

[12] As defined by Aragonès et al. (2007)
[13] That equates *cost of reneging* to the *gain from reneging*.

be equal to d in equilibrium, and hence -

$$V_{GG} - V_{BG} = d$$

$\Rightarrow (v_{GG} + \delta v_{GG} + \delta^2 v_{GG} + \delta^3 v_{GG} + ...\infty) - (v_{GG} + \delta v_{BG} + \delta^2 v_{BG} + \delta^3 v_{GG} + ...\infty) = d$

$\Rightarrow \delta(v_{GG} - v_{BG}) + \delta^2(v_{GG} - v_{BG}) = d$

Putting the respective values of $v_{GG}$ and $v_{BG}$, we find

$\Rightarrow \delta[(1 - d + d^2/3)] + \delta^2[(1 - d + d^2/3)] = 1$

$\Rightarrow \delta(1 + \delta)[(1 - d + d^2/3)] = 1$

The same result is derived when we calculate the equilibrium condition with the opponent now being *bad*. Hence, for k = 1, we can write the following equilibrium condition to be one as -

$$V_{GG} - V_{BG} = V_{GB} - V_{BB} = d$$

$$\Rightarrow \delta(1 + \delta)[(1 - d + d^2/3)] = 1 \qquad (18)$$

PROPOSITION 3. *The strategies just described constitute an equilibrium. The promises believed and fulfilled in equilibrium with linear utility functions are those within d\*(δ) of the candidates' ideal points, where (Aragonès et al., 2007)*

$$d_{k1}^*(\delta) = \begin{cases} 0 \text{ if } 0 \leq \delta < \frac{3}{5} \\ \\ \frac{3}{2}(1 - \sqrt{(\frac{4}{3\delta(1+\delta)} - \frac{1}{3})}) \text{ if } \frac{3}{5} \leq \delta \leq 1 \end{cases} \qquad (19)$$

### Results for k=2 and k=3

Similarly we find the following equilibrium conditions and their respective *d\** for k=2 and 3 respectively. They are as follows -

$$V_{GG} - V_{BG} = V_{GB} - V_{BB} = d$$

$$\Rightarrow \delta(1+\delta+\delta^2)[(1-d+d^2/3)] = 1 \qquad (20)$$

and,

$$V_{GG} - V_{BG} = V_{GB} - V_{BB} = d$$

$$\Rightarrow \delta(1+\delta+\delta^2+\delta^3)[(1-d+d^2/3)] = 1 \qquad (21)$$

PROPOSITION 3. *The promises believed and fulfilled in equilibrium (for k =2 and 3, rspectively) with linear utility functions are those within d\*(δ) of the candidates' ideal points, where*

$$d_{k2}^*(\delta) = \begin{cases} 0 \text{ if } 0 \leq \delta < \frac{11}{20} \\ \\ \frac{3}{2}(1-\sqrt{(\frac{4}{3\delta(1+\delta+\delta^2)} - \frac{1}{3})}) \text{ if } \frac{11}{20} \leq \delta \leq 1 \end{cases} \qquad (22)$$

$$d_{k3}^*(\delta) = \begin{cases} 0 \text{ if } 0 \leq \delta < \frac{3}{5} \\ \\ \frac{3}{2}(1-\sqrt{(\frac{4}{3\delta(1+\delta+\delta^2+\delta^3)} - \frac{1}{3})}) \text{ if } \frac{3}{5} \leq \delta \leq 1 \end{cases} \qquad (23)$$

## Mechanics underlying the Results

From Fig 2.2 below, we find that for an appropriate value of the discount factor $\delta = 0.6$, the values of $d_{k1}^*(\delta)$, $d_{k2}^*(\delta)$, $d_{k3}^*(\delta)$, and $d_{kA}^*(\delta)$ [14] are in ascending order to each other respectively. That is, for $\delta = 0.6$, we have -

$$d_{k1}^*(\delta = 0.6) < d_{k1}^*(\delta = 0.6) < d_{k1}^*(\delta = 0.6) < d_{kA}^*(\delta = 0.6) \qquad (24)$$

From equation (16) we find that for an appropriate value of $\delta$, the $d^*$s for k = 1,2, and 3 approaches the Aragonès et al. (2007) result from the *left*. This result can be intuitively thought of as follows - For lower values of punishment periods by voters, the future expected pay-offs of a candidate reneging is *higher* in equilibrium. This reasoning can be extended to note that this would lower the cost of reneging[15]. And since we are trying to find out the maximal credible promise in equilibrium, given by -

$$d \leq C^S$$

---
[14] Here $d_{kA}^*(\delta)$ denotes the equilibrium $d^*$ under Aragonès et al. (2007)

Lower the value of the cost of reneging would equivalently mean a lower $d^*$ in equilibrium, and hence the result.

In this section we will delve into the exact mathematical derivation of the equations (12) and (13), and how these equations lead us to the respective $d^*$s of equations (14) and (15).

**Understanding the result for k =2**

First, we try to find the equilibrium condition that would be used to calculate the equilibrium $d^*$ when the opponent is *good*. Applying the same logic as the one used to calculate for the condition for k = 1, we have for k = 2 -

$$(v_{GG}+\delta v_{GG}+\delta^2 v_{GG}+\delta^3 v_{GG}+\delta^4 v_{GG}+...\infty) - (v_{GG}+\delta v_{BG}+\delta^2 v_{BG}+\delta^3 v_{BG}+\delta^4 v_{GG}+...\infty) = d$$

$$\Rightarrow (\delta + \delta^2 + \delta^3)(v_{GG} - v_{BG}) = d$$

$$\Rightarrow \delta(1 + \delta + \delta^2)(1 - d + d^2/3) = 1 \tag{25}$$

| Time | 0 | 1 | 2 | 3 | 4 |
|---|---|---|---|---|---|
| Candidate | G | G (B) | G (B) | G (B) | G |
| Opponent | B | B | B | G | G |

Now, when the candidate's opponent is *bad*, we refer to the table above and notice that the candidate has one of the two periods (either period 1 or period 2) to decide whether to renege or not. We take up the first case, where the candidate decides whether to renege (or not) in period 1, when the opponent is *bad*[16] -

$$\Rightarrow (v_{GB}+\delta v_{GB}+\delta^2 v_{GB}+\delta^3 v_{GG}+\delta^4 v_{GG}+...\infty)-(v_{GB}+\delta v_{BB}+\delta^2 v_{BB}+\delta^3 v_{BG}+\delta^4 v_{GG}+...\infty) = d$$

$$\Rightarrow \delta(v_{GB} - v_{BB}) + \delta^2(v_{GB} - v_{BB}) + \delta^3(v_{GG} - v_{BG}) = d$$

$$\Rightarrow \delta(1 + \delta + \delta^2)(1 - d + d^2/3) = 1 \tag{26}$$

---
[15]Since the cost of reneging ($C^S$) is equivalent to $V_G$ - $V_B$ (irrespective of the opponent's type). The rise in $V_B$ would lower the cost of reneging.

Thus, from eq. (17) and (18), we find that the equilibrium condition (and hence $d^*$) is equal for a *good* candidate irrespective of whether his opponent is *good* or a *bad* candidate. So, the Aragonès et al. (2007) result holds for k = 1. The same result holds if the candidate decides to renege in period 2 instead of period 1.

**Understanding the result for k =3**

As in the previous section, we are going to solve the problem from the candidate's perspective - once, by allowing the opponent to be *good* and in the other case, to be *bad*. So, let us begin with the case where his opponent is *good*.

| Time      | 0 | 1        | 2        | 3        | 4        |
|-----------|---|----------|----------|----------|----------|
| Candidate | G | G<br>(B) | G<br>(B) | G<br>(B) | G<br>(B) |
| Opponent  | G | G        | G        | G        | G        |

As seen from the table above, both the candidates' enter the game as *good* where the candidate (winner) decides to renege or not. Then, his pay-off from *not reneging* can be shown to be equal to -

$$(v_{GG} + \delta v_{GG} + \delta^2 v_{GG} + \delta^3 v_{GG} + \delta^4 v_{GG} + ...\infty)$$

Similarly, the pay-offs from *reneging* can be shown to be equal to -

$$(v_{GG} + \delta v_{BG} + \delta^2 v_{BG} + \delta^3 v_{BG} + \delta^4 v_{BG} + \delta^5 v_{GG} + ...\infty)$$

---

[16] As shown in Aragonès et al. (2007): $(v_{GG} - v_{BG}) = (v_{GB} - v_{BB})$

Then the cost of reneging is equal to -

$$C^S = (\text{Not Renege}) - (\text{Renege}) = (V_{GG} - V_{BG})$$

Then, in equilibrium we have -

$$\text{Cost of Reneging} = \text{Gain from Reneging}$$

$\Rightarrow (v_{GG} + \delta v_{GG} + \delta^2 v_{GG} + \delta^3 v_{GG} + \delta^4 v_{GG} + ...\infty) - (v_{GG} + \delta v_{BG} + \delta^2 v_{BG} + \delta^3 v_{BG} + \delta^4 v_{BG} + \delta^5 v_{GG} + ...\infty) = d$

$\Rightarrow \delta(v_{GG} - v_{BG}) + \delta^2(v_{GG} - v_{BG}) + \delta^3(v_{GG} - v_{BG}) + \delta^4(v_{GG} - v_{BG}) = d$

$\Rightarrow (\delta + \delta^2 + \delta^3 + \delta^4)(v_{GG} - v_{BG}) = d$

The result finally boils down to -

$$\delta(1 + \delta + \delta^2 + \delta^3)(1 - d + d^2/3) = 1 \qquad (27)$$

| Time | 0 | 1 | 2 | 3 | 4 | 5 |
|---|---|---|---|---|---|---|
| Candidate | G | G (B) | G (B) | G (B) | G | G |
| Opponent | B | B | B | B | G | G |

We now move to the case with the opponent being *bad*. In such a scenario, there will exist three (3) different cases. The three different cases exist because of the fact that the candidate (winning) will have three distinct time points over which he can decide whether to renege or not. The table above illustrates the first case where at time t = 0, the candidate (winner) and his opponent begin with the types *good* and *bad* respectively. As we see, the opponent remains *bad* on t = 1, t = 2 and t = 3, after which his reputation is restored to that of *good*. So, the candidate has three (3) distinct points of time, they being - t = 1, t = 2 and t = 3 - to decide whether to renege or not.

First we consider the case where the candidate is deciding to renege at t = 1 (as shown in

| Time      | 0 | 1 | 2 | 3 | 4 | 5 |
|-----------|---|---|---|---|---|---|
| Candidate | G | B | B | B | B | G |
| Opponent  | B | B | B | B | G | G |

the table above). Then applying the same methodology as the one used above, we have -

$$(v_{GB}+\delta v_{GB}+\delta^2 v_{GB}+\delta^3 v_{GB}+\delta^4 v_{GG}+...\infty)-(v_{GB}+\delta v_{BB}+\delta^2 v_{BB}+\delta^3 v_{BB}+\delta^4 v_{BG}+\delta^5 v_{GG}+...\infty) = d$$

$$\Rightarrow \delta(v_{GB}-v_{BB})+\delta^2(v_{GB}-v_{BB})+\delta^3(v_{GB}-v_{BB})+\delta^4(v_{GG}-v_{BG}) = d$$

The result finally becomes -

$$\delta(1+\delta+\delta^2+\delta^3)(1-d+d^2/3) = 1 \qquad (28)$$

We find that the equilibrium condition, as shown by equations (19) and (20) respectively, are equal. This result necessarily points to the fact that the $d^*$ would also be the same. In fact, we will further show that for the remaining two (2) others cases here, the equilibrium condition would exactly be equal to the one in equation (19).

| Time      | 0 | 1 | 2 | 3 | 4 | 5 |
|-----------|---|---|---|---|---|---|
| Candidate | G | G | B | B | B | B |
| Opponent  | B | B | B | B | G | G |

Moving on to the case where the candidate (winning) is deciding whether to renege or not at t = 2 (as shown in the table above), we have -

$$(v_{GB}+\delta v_{GB}+\delta^2 v_{GB}+\delta^3 v_{GG}+\delta^4 v_{GG}+...\infty)-(v_{GB}+\delta v_{BB}+\delta^2 v_{BB}+\delta^3 v_{BG}+\delta^4 v_{BG}+\delta^5 v_{GG}+...\infty) = d$$

$$\Rightarrow \delta(v_{GB}-v_{BB})+\delta^2(v_{GB}-v_{BB})+\delta^3(v_{GG}-v_{BG})+\delta^4(v_{GG}-v_{BG}) = d$$

The result finally becomes -

$$\delta(1+\delta+\delta^2+\delta^3)(1-d+d^2/3) = 1 \qquad (29)$$

And, to the final case where the candidate (winning) is deciding whether to renege or not at

| Time      | 0 | 1 | 2 | 3 | 4 | 5 | 6 |
|-----------|---|---|---|---|---|---|---|
| Candidate | G | G | G | B | B | B | B |
| Opponent  | B | B | B | B | G | G | G |

t = 3 (as shown in the table above), we have -

$$(v_{GB}+\delta v_{GB}+\delta^2 v_{GG}+\delta^3 v_{GG}+\delta^4 v_{GG}+...\infty)-(v_{GB}+\delta v_{BB}+\delta^2 v_{BG}+\delta^3 v_{BG}+\delta^4 v_{BG}+\delta^5 v_{GG}+...\infty) = d$$

$$\Rightarrow \delta(v_{GB}-v_{BB})+\delta^2(v_{GG}-v_{BG})+\delta^3(v_{GG}-v_{BG})+\delta^4(v_{GG}-v_{BG})=d$$

The result finally becomes -

$$\delta(1+\delta+\delta^2+\delta^3)(1-d+d^2/3)=1 \qquad (30)$$

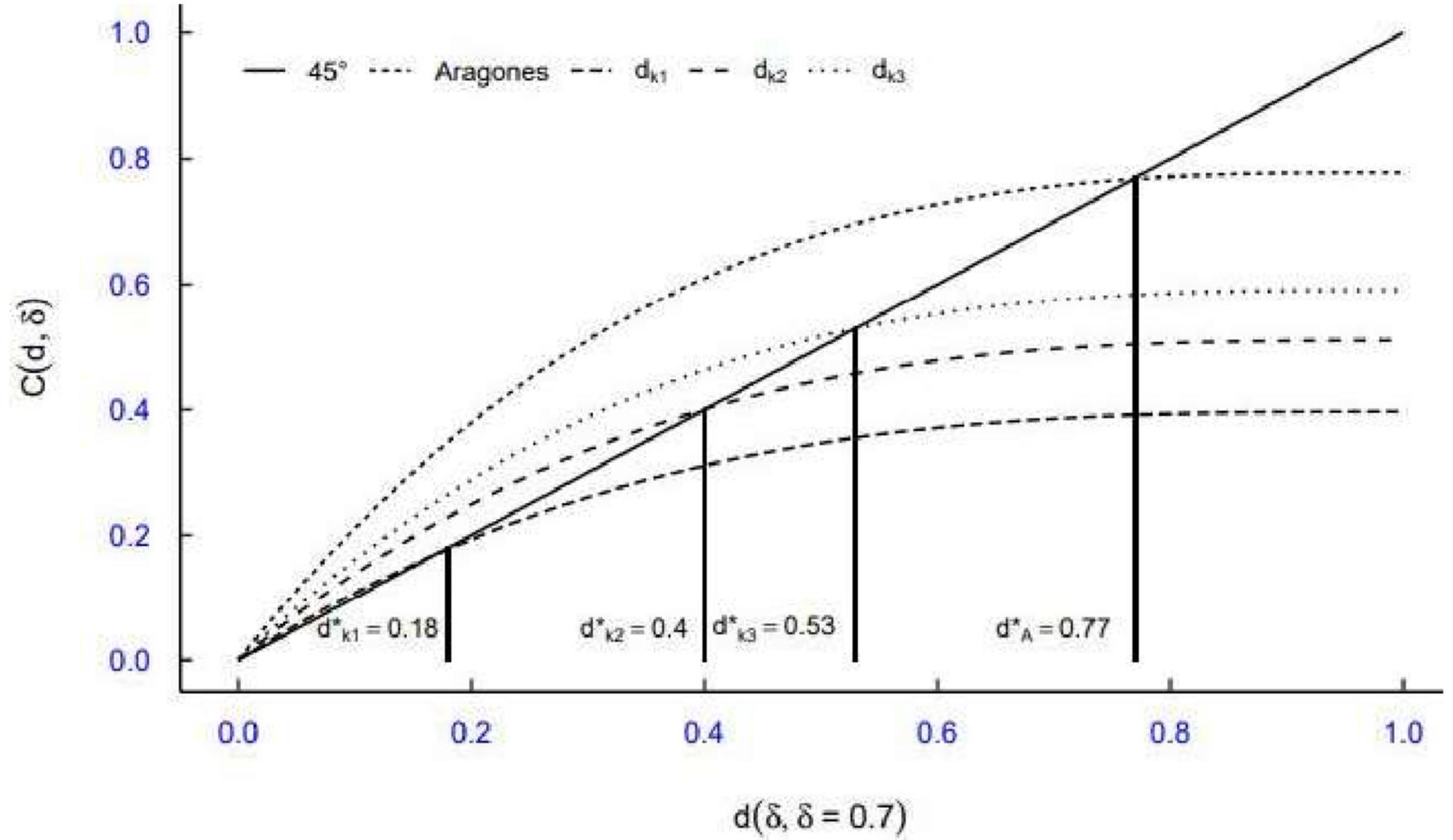

Figure 2: Comparative analysis of $d_A^*$, $d_{k1}^*$, $d_{k2}^*$, and $d_{k3}^*$ for $\delta = 0.7$

As shown by Figure 2.2 below, we find graphical evidence of our result - with each increase in the period of punishment, the equilibrium value of the maximal credible promise *converges* to the Aragonès et al. (2007)'s result as the period of punishment tends to infinity. Here this is can be seen that for an appropriate value of the discount factor ($\delta = 0.7$) - the values of the respective $d^*$s are arranged in the ascending order starting with $d^*_{k1}$, then $d^*_{k2}$, $d^*_{k3}$ and $d^*_A$ respectively. Mathematically, this can be shown to be as -

$$d^*_{k1} < d^*_{k2} < d^*_{k3} < d^*_A$$

for $\delta = 0.7$

## Implication of the Results

The strategy of incorporating different periods of punishment can be thought of as a pertinent tool for *naive* voters to achieve *equivalent* (not equal) results as those achieved by voters who are able to behave *strategically*. This inference can be drawn from the fact that the new $d^*$s under both limited punishment and non - naive voting behavior is *lower* than the result put forth by Aragonès et al. (2007), ceteris paribus. Thus, more *forgiving* behavior on part of the voters reduces the incentive-compatible promises made by each candidate - the incentive to sell short (and maintain one's own reputation) finds higher precedence than that found in the benchmark model. Interestingly, laboratory experiments on social identity and candidate credibility has shown that voters (the principals) belonging to the same ethnic or racial background as the candidate(s) are more prone to treat the latter with laxer standards of re-election, which in turn induces the same candidates (agents) to lower their effort and in the long-run, competence (Duell and Landa (2021) , Landa and Duell (2015)).